\documentclass[10pt,letterpaper]{article}
\usepackage[top=0.85in,left=1.75in,footskip=0.75in]{geometry}

\usepackage[utf8]{inputenc}

\usepackage{textcomp,marvosym}

\usepackage{fixltx2e}

\usepackage{amsmath,amssymb}

\usepackage{cite}

\usepackage{nameref,hyperref}

\usepackage{microtype}
\DisableLigatures[f]{encoding = *, family = * }

\usepackage{rotating}

\raggedright
\usepackage[aboveskip=1pt,labelfont=bf,labelsep=period,justification=raggedright,singlelinecheck=off]{caption}

\makeatletter
\renewcommand{\@biblabel}[1]{\quad#1.}
\makeatother

\date{}

\usepackage{graphicx}
\begin{document}
\vspace*{0.35in}

% Title must be 150 characters or less
\begin{flushleft} {\Large \bf Balance of interactions determines
    optimal survival in multi-species communities}
%Effect of diversity in interactions on the survival of multi-scpecies communities}
% Optimal Survival in Model Ecosystems}} 
% Another possible title could be : "Diversity in interactions leads
% to complex spatiotemporal behavior in multiscpecies communities."
\newline
% Insert Author names, affiliations and corresponding author email.
\\
{\bf Anshul Choudhary and
Sudeshna Sinha}
%\textsuperscript{1}
%Name3 Surname\textsuperscript{2,\textcurrency a},
%Name4 Surname\textsuperscript{2,\ddag},
%Name5 Surname\textsuperscript{2,\ddag},
%Name6 Surname\textsuperscript{2,\Yinyang},
%Name7 Surname\textsuperscript{3,*,\Yinyang}
\\
%{1} 

Department of Physical Sciences, Indian Institute of Science Education and Research (IISER) Mohali, Knowledge City, SAS Nagar, Sector 81, Manauli PO 140 306, Punjab, India.
%\\
%\bf{2} Affiliation Dept/Program/Center, Institution Name, City, State, Country
%\\
%\bf{3} Affiliation Dept/Program/Center, Institution Name, City, State, Country
%\\

% Insert additional author notes using the symbols described below. Insert symbol callouts after author names as necessary.
% 
% Remove or comment out the author notes below if they aren't used.
%
% Primary Equal Contribution Note
%\Yinyang These authors contributed equally to this work.

% Additional Equal Contribution Note
%\ddag These authors also contributed equally to this work.

%\textcurrency a Insert current address of first author with an address update
% \textcurrency b Insert current address of second author with an address update
% \textcurrency c Insert current address of third author with an address update


% Group/Consortium Author Note
%\textpilcrow Insert Collaborative Author line here

* E-mail of corresponding author:  sudeshna@iisermohali.ac.in
\end{flushleft}
\section*{Abstract}

We consider a multi-species community modelled as a complex network of
populations, where the links are given by a random asymmetric
connectivity matrix $J$, with fraction $1-C$ of zero entries, where
$C$ reflects the over-all connectivity of the system. The non-zero
elements of $J$ are drawn from a gaussian distribution with mean $\mu$
and standard deviation $\sigma$. The signs of the elements $J_{ij}$
reflect the nature of density-dependent interactions, such as
predatory-prey, mutualism or competition, and their magnitudes reflect
the strength of the interaction. In this study we try to uncover the
broad features of the interspecies interactions that determine the
%%change
global robustness of this network, as indicated by
%.  As an indicator of global stability, we first calculate 
the average number of active nodes (i.e. non-extinct species) in the
network, and the total population, reflecting the biomass yield. 
We find that the network transitions from a completely extinct system
to one where all nodes are active, as the mean interaction strength
goes from negative to positive, with the transition getting sharper
for increasing $C$ and decreasing $\sigma$.
%%end_change
We also find that the total population, displays distinct
non-monotonic scaling behaviour with respect to the product $\mu C$,
implying that survival is dependent not merely on the number of links,
but rather on the combination of the sparseness of the connectivity
matrix and the net interaction strength. Interestingly, in an
intermediate window of positive $\mu C$, the total population is
maximal, indicating that too little or too much positive interactions
is detrimental to survival. Rather, the total population levels are
optimal when the network has intermediate net positive connection
strengths.  At the local level we observe marked qualitative changes
in dynamical patterns, ranging from anti-phase clusters of period $2$
cycles and chaotic bands, to fixed points, under the variation of mean
$\mu$ of the interaction strengths. We also study the correlation
between synchronization and survival, and find that synchronization
does not necessarily lead to extinction.
%%change
Lastly, we propose an effective low dimensional map to
capture the behavior of the entire network, and this provides a broad
understanding of the interplay of the local dynamical patterns and the
global robustness trends in the network.
%%end_change

%\linenumbers

\section*{Introduction}

Complex networks provides a common framework to address a vast range
of phenomena in large interactive systems
%like neural networks, ecosystems, coupled oscillators, world-wide-web, etc
\cite{newman}. The use of network theory in studying the stability and
dynamics of model ecosystems started with the landmark paper of Robert
May\cite{May72} and the success and effectiveness of such inquiry can
be gauged from the fact that even today most studies in theoretical
ecology heavily rely on the network framework \cite{general}. Our
current understanding of the stability of an ecological network hinges
around two key aspects: {\em interaction topology} and {\em nature of
  interactions}.
%The first factor reflects the existence of well
%defined network topologies, like nestedness and modular structure. 
Now, there are wide ranging situations where one either does not have
sufficient information on the exact underlying topology, or one finds
that the web of interactions essentially appears to be a random
network. In such cases, the interactions are modelled by random
connectivity matrices, and the broad nature of interactions is our
only guiding principle in analyzing the dynamics and survival
properties of the complex system.

%Now, most studies focus on statistical aspects and do not focus on the
%emergent dynamical complexity of the collective behavior. 

In this study, we are going to explore the effect of the balance of
different kinds of interactions in a multi-scpecies community on the
collective dynamical behaviour of the network. 
%%change
Our focus will be on the {\em global robustness} of the system, as
exemplified by the total population of all species
\cite{stability1}. It is evident that the total population relects the
global state of the network effectively, while being sensitive to the
underlying dynamics at the local species level as well. So here we
will explore how {\em diversity in interactions influence the emergent
  dynamics}, and the {\em relation of these dynamical patterns to
  survival of populations}, lending yet another perspective to the
stability-diversity debate.
%%end_change

Recently, Mougi \& Kondoh\cite{stability3} studied the interesting
effects of diversity in interaction types on the stability of an
ecological community and they found that diversity is a key element in
determining stability and biodiversity. However their results are
based on linear stability analysis for small perturbations about local
equilibria, and they do not give the relationship between survival and
the emergent dynamical patterns. In this context, our study provides a
complementary exploration of the global survival features of such
systems \cite{sinha} and also relates it to dynamical behavior of the
constituent populations.

\section*{Model}

%%Change
The model we consider here is inspired by the earlier theoretical
studies conducted by Robert May \cite{May72,May73}.  However, we would
like to mention here that unlike most studies regarding stability
\cite{stability2,stability3} which assumes the existence of
time-independent population densities when system reaches steady
state, {\em we consider the more general condition where the
  attracting state can have complex temporal behavior}, rather than a
fixed point solution \cite{sinha}. The principal motivation for this
approach to the question of stability is its wider relevance and
broader applicability. Further, rather than local stability about an
equilibrium, we will focus on a different set of {\em global
  quantifiers of robustness and survival in the complex network}.
%%end_change

Specifically, in this work we consider a prototypical map, the Ricker
(Exponential) Map, modelling population growth of species with
non-overlapping generations:
\begin{equation} 
f(x) = x \ e^{r(1-x)}
\end{equation}
with $r$ interpreted as an intrinsic growth rate. 
%In this work $r=4$, namely the local dynamics is chaotic. 

We then consider the evolution of $N$ such interacting populations
given as:
\begin{equation} 
x_{i}(n+1) = f({x}_{i}(n)) +\frac{1}{N}\sum_{j}J_{ij}x_{j}(n)
\label{system} 
\end{equation}
where $i = 1, \dots N$, and the connectivity or community matrix, $J$
represents how species are mutually interacting.
Further we consider that $x_{i} (n+1) = 0$ if $x_i(n+1) < x_{threshold}$,
%\begin{equation} 
%f(x) = \left\{ \begin{array}{rl} x e^{r(1-x)} & \ \ \ \ \mbox{
%	        if $x > x_{threshold}$,} \\ 0 & \ \ \ \ \mbox{ otherwise,} \end{arra%y} \right.
%\label{map} 
%\end{equation}			
where $x_{threshold}$ is the minimum population density below which
population cannot sustain on their own and therefore becomes extinct,
namely the Allee Effect \cite{allee}. So we have a system
(Eqn.~\ref{system}) with well mixed populations at the nodes that
display chaotic dynamics, and are extinction prone due to the
population threshold when uncoupled. {\em So clearly, a persistent
  community can only be sustained through suitable interactions among
  the species.}

The connectivity matrix $J$ is a random asymmetric matrix, with
fraction $1-C$ of zero entries, where $C$ reflects the over-all
connectivity of the system. The non-zero elements are drawn from a
gaussian distribution, $\frac{1}{\sqrt{2\pi{\sigma}^{2}}}
e^{\frac{-(x-\mu)^{2}}{2{\sigma}^{2}}}$, with mean $\mu$ and standard
deviation $\sigma$.

The signs of the elements $J_{ij}$ of the connectivity matrix $J$
reflects the nature of  density-dependent interactions. In general, neutral
interactions are reflected by zero matrix elements. Interactions that
reduce the population at a node, for instance through parasitism,
grazing, and predation, will be reflected by a negative sign, while
interactions that benefit a species, for instance by provding refuge
from physical stress, predation or competition, will bear a positive
sign. 

So then, when we have mutualism or symbiosm, both $J_{ij}$ and
$J_{ji}$ are positive. When we have competition or antagonistic
interactions, both $J_{ij}$ and $J_{ji}$ are negative. When the effect
of one species on the other is positive, but neutral the other way
round, we have commensalism, reflected by $J_{ij} > 0$ and $J_{ji} =
0$. Similarly, amensalism is reflected by $J_{ij} < 0$ and $J_{ji} =
0$. General predator-prey interactions are captured by $J_{ij}$ and
$J_{ji}$ having different signs.

In general, positive interactions or facilitative interactions between
species that benefit the growth of the species, give rise to more
positive mean $\mu$. Namely, high positive $\mu$ implies the dominance
of mutualism in the ecosystem where as $\mu \sim 0$ would imply
balance of different kinds of interactions in the system. The standard
deviations, $\sigma$ on the other hand controls the degree of
variability in the strength of these interactions.  Finally,
connectedness, $C$ is another important factor that tell us how many
species are interacting with each other. Namely, it reflects the
fraction of neutral interactions in the network.  {\em The main aim of
  this study is to understand the relation between broad features of
  the interaction matrix and the collective dynamics of the system,
  and then go on to link this to the local and global survival in the
  system.}
  
%%change
\section*{Methods}

At the outset, we present our tools and describe the measures for
analyzing the survival properties of the system. To gauge the
robustness of the system, we first calculate the {\em number of active
  nodes}, namely the number of non-extinct species with non-zero
population, after transience. This quantity is then averaged over a
period of time $T$ and further averaged over $N_{ic}$ different
initial conditions. We denote this averaged number of non-extinct
nodes by $\langle N_{active} \rangle$, and it is defined as:
\begin{align}
 \langle N_{active} \rangle &= \frac{1}{N_{ic} T} \sum_{N_{ic}}\sum_{T} \ \left\{ \sum_{i=1}^{N}\phi_{i} (t) \right\} \ \ \ {\rm where},\\
 \phi_{i} (t) &= \left\{ \begin{array}{rl} 1 & \ \ \ \ \mbox{
	        if $x_i (t) > x_{threshold}$,} \\ 0 & \ \ \ \ \mbox{ otherwise,} \end{array} \right.
\label{nactive}
\end{align}

The next important measure of global survival is the {\em total
  population} $\Sigma_{i=1}^N x_i$, of the system, reflecting the
biomass yield in a multi-species community. This quantity, averaged
over a period of time $T$ and over $N_{ic}$ different initial
realizations, is denoted by $\langle x_{total} \rangle$, and
mathematically expressed as:
\begin{equation}
\langle x_{total} \rangle = \frac{1}{N_{ic} T} \sum_{N_{ic}}\sum_{T} \ \left\{ \sum_{i=1}^{N}x_{i}(t) \right\}
\label{xtotal}
\end{equation}

{\em Synchronization Order Parameter :}\\

In order to probe collective patterns in the network, we studied the
level of synchronization that emerges in the system. To quantify the
degree of synchronization we have employed two different order
parameters.

(i) $Z_{sync}$: Here we measure synchronization error as the mean
square deviation of the local state of the nodes, averaged over time
$T$ (after transience) and over $N_{ic}$ different realizations
\cite{msd1,msd2,msd3}, mathematically expressed as :
\begin{equation}
 Z_{sync} =  \frac{1}{N_{ic} T} \sum_{N_{ic}}\sum_{T} \ \left\{ \frac{1}{N} \sum_{i=1}^{N} \left[ x_i (t) - \langle x (t) \rangle \right]^2 \right\}
\label{zsync}
\end{equation}
When it goes to zero, this measure reflects {\em complete
  synchronization} in the system.

(ii) $Z_{phase}$: This is a phase order parameter that reflects the
degree of variation in the phases of the local dynamics at the
nodes. Specifically, it is a measure of the fraction of nodes in the
largest phase cluster, averaged over time $T$ and over different
network realizations $N_{ic}$. When $Z_{phase} = 1$, it implies that
the entire system is phase synchronized (though not necessarily in
complete synchronization).

For the specific case of the local dynamics being a $2$-cycle or being
in a period $2$ chaotic band (which is observed in this system over a
large parameter range) $Z_{phase}$ then is the supremum of the
quantity $$\frac{1}{N_{ic} T} \sum_{N_{ic}}\sum_{T} \ \left\{
  \frac{1}{N} \sum_{i=1}^{N}\varphi_{i} (t) \right\}$$ where
$\varphi_{i} (t)$ is $1$ if $x_i(t)$ lies in the specified band, and
$0$ otherwise.
%\begin{align}
% Z_{bands} &= \frac{1}{N_{ic} T} \sum_{N_{ic}}\sum_{T} \frac{1}{N}\sum_{i=1}^{N%}\varphi_{i} \ \ \ where, \\
% \varphi_{i} &= \left\{ \begin{array}{rl} 1 & \ \ \ \ \mbox{
%	        if $x(t)$ belongs to the specified band,} \\ 0 & \ \ \ \ \mbox{ %otherwise,} \end{array} \right.
%\end{align}

So these measures provide complementary information about the
synchrony in phase and amplitude of the dynamics of the local
constituents of the network.

\medskip
{\em System Parameters: }\\

In this work parameter $r=4$ in Eq. 1, namely the local map is in the
chaotic regime, and the threshold value $x_{threshold} = 0.0001$. 
%%Correct thrshold value?
All results reported here are robust with respect to small variations
around these values. The survival and synchronization measures were
calculated by averaging over $100$ random initial conditions, i.e.
$N_{ic}=100$ in the equations above. The system sizes ranged from
$100\le N \le 800$, with connectedness $0\le C \le1$ and standard
deviation $0.1 \le \sigma \le0.5$ in the connectivity matrix
$J$. Further, to explore the effect of the mean interaction strength
$\mu$ on the dynamics, which is a focus of our work here, we
investigated the range: $-1 \le \mu \le 1$. Note that several earlier
studies have been confined to the balanced situation, where $\mu =
0$. In the sections below, we present the principal observations from
our extensive simulations over this wide-ranging window of parameters.

%%end_change

\bigskip
\section*{Results and Analysis}

\subsection*{Survival in the Network}

We first calculate the average number of active nodes (namely the
average number of non-extinct species) $\langle N_{active} \rangle$,
as a function of the mean interaction strength $\mu$ of the
connectivity matrix $J$. As evident from the results displayed in
Fig. 1, the average number of active nodes $\langle N_{active}
\rangle$ in the network rises sharply as a function of mean
interaction strength $\mu$ around $\mu \sim 0$. When the mean
interaction strength is quite negative, the number of active nodes
goes to zero, i.e. the entire system is driven to extinction. For
positive $\mu$ all nodes in the network are non-zero, i.e. no species
goes extinct. The connectedness $C$ and the variability of interaction
strengths $\sigma$ then does not affect the number of active nodes in
the network when the network is far from balanced, namely considerably
positive mean interaction strengths yield $\langle N_{active} \rangle
= N$, while considerably negative interactions results in $\langle
N_{active} \rangle = 0$, irrespective of $C$ and $\sigma$.
The transition from complete extinction to a completely active network
is sharper for systems with low variability in interaction strengths
(i.e. low $\sigma$), and for systems with higher connectedness
(i.e. high $C$).

%% Para below replaced by scaling results
% Interestingly, when the network is close to the balanced situation,
% the number of active nodes depends crucially on the connectedness $C$
% and variability $\sigma$ of the interaction strengths. When the mean
% interaction strength is slightly negative, the number of active nodes
% increases as $C$ decreases and $\sigma$ increases. On the other hand,
% when the mean interaction strength is slightly positive, the number of
% active nodes increases as $C$ increases and $\sigma$ decreases, namely
% the trend is completely reversed. This change in the dependence of
% network activity on $C$ and $\sigma$ occurs at $\mu \sim 0$.
% 
% Notice that these broad trends are quite distinct from earlier
% stability results obtained for strictly balanced networks
% \cite{May72,sinha}. So one can infer that when the assumed balance
% between interactions breaks down, there is a very significant
% departure from the May-Wigner scenario. This is very relevant, as real
% ecosystems are unlikely to have a perfect balance of interactions. So
% it is important to explore the whole range of interaction strengths in
% order to understand the effects of the imbalance of interactions
% \cite{balanced} on the behavior of ecosystem.

%%changes 
To gain further quantitative understanding of the nature of this
transition, we explore the scaling behaviour near the transition, and
discover that the average number of active sites scales with respect to $C$ as:
%\begin{align}
% \langle N_{active} \rangle &\sim \Theta((\mu-\mu_{c})C^{\alpha}) \ \ \ {\rm and},\\
%			    &\sim \Omega((\mu-\mu_{c})\sigma^{\beta}) 
%\end{align}
\begin{equation}
 \langle N_{active} \rangle \ \sim \ \Theta((\mu-\mu_{c})C^{\alpha})
\end{equation}
Further the number of active sites scales with respect to $\sigma$ as:
\begin{equation}
 \langle N_{active} \rangle \ \sim \ \Omega((\mu-\mu_{c})\sigma^{\beta}) 
\end{equation}
Here, $\alpha$ and $\beta$ are appropriate critical exponents for
scaling functions $\Theta$ and $\Omega$ respectively.  A good data
collapse, shown in Fig. 1 (insets), is obtained for $\mu_c = 0$, with
$\alpha = 0.45 \pm 0.02$ and $\beta = 1$. These scaling relations
suggest that a transition from complete extinction to a fully active
ecosystem occurs around $\mu = 0$, namely around the state of
balanced interaction strengths or completely neutral interactions
\cite{balanced}. We also performed finite size scaling with respect to
system size $N$, and found the simple scaling: $\langle N_{active}
\rangle \sim N f(\sigma,\mu, C)$, implying that the active fraction
$\langle N_{active} \rangle/N$ is independent of $N$.

%, with the nature of the transition being governed by an interplay of $\mu$, $C$ and $\sigma$

% \begin{equation}
%  \abs{\mu-\mu_{c}} \sim \frac{1}{C^{0.45}\sigma}
% \end{equation}

\medskip

The next important measure of global survival is the average total
population $\langle x_{total} \rangle$ of the system, reflecting the
biomass yield in a multi-species community. The variation of $\langle
x_{total} \rangle$, as a function of mean $\mu$ of the interaction
strengths and connectedness $C$ of the interaction matrix $J$, is
shown in Figs. 2-3. It is clear that for $C=0$, i.e, when there are no
interactions, the local extinctions accumulate, eventually leading to
mass extinction.

When interactions are present, different global scenarios emerge with
respect to varying mean $\mu$ and connectedness $C$ of matrix $J$. For
fixed $C$, the total population increases sharply with increasing
$\mu$, namely with increasing net positive interactions, around $\mu
\sim 0$. 
%Note that $\mu = 0$ is the situation where, on an average,
%the net strengths of the positive interaction balance those of the
%negative interactions \cite{balanced}. 
So we find that for networks close to the balanced situation, we have
enhanced population densities indicating greater survival, for
increasing net positive interactions.

Further, there exists an interval of mean $\mu$, around $\mu \sim 0$,
where the average total population always increases with the increase
in the number of interactions among species, namely increasing
$C$. This would imply that connectivity always enhances survival of
the system here.
% Such monotonic dependence of survival on $C$ has been observed in earlier studies \cite{sinha}.
However, when mean $\mu$ is smaller, or larger, than the above
interval, one finds that at {\em
%%change 
 %intermediate values of connectedness, the population density is the
  intermediate values of connectedness, the population is the
  largest}(cf. Fig.2). Namely, a mix of neutral interactions along-side other
%%change
  % interactions is most conducive to enhanced population density.
interactions is most conducive to enhanced population yeild.

  There is a critical negative mean, $\mu_{c}^{extinction}$ (where
  $\mu_c^{extinction}$ is a function of $C$) for which the local
  species experience severe loss of population leading to
  global extinction. There is also a critical positive mean,
  $\mu_{c}^{explosion}$, where $\mu_c^{explosion} C \sim 1$, such that
  for mean $\mu > \mu_{c}^{explosion}$ the nodes experience unbounded
  and explosive growth, destabilizing the whole network. We consider
  $\mu < \mu_c^{explosion}$ in our study.

  Interestingly, we uncovered a scaling pattern between the total
  population and characteristics of the connectivity matrix. The data
  collapse of the population onto a single {\em non-monotonic} curve
  in Fig.~3B reveals a scaling relation between total population and
  the product of the mean interaction strength, $\mu$ and
  connectedness, $C$ of the network. This implies that the most
  relevant quantity in understanding the global behaviour of the
  network is $\mu C$, rather than $\mu$ and $C$ alone. So clearly,
  survival is dependent not merely on the number of links, but on the
  combination of the sparseness of the connectivity matrix and the net
  interaction strength. For instance, fewer interactions (i.e. low
  $C$) tends to decrease the population, but this effect may be
  compensated by more positive interactions, i.e. higher $\mu$. More
  importantly, the existence of an intermediate window of positive
  $\mu C$ where the total population is maximum indicates that too
  little or too much positive interactions is detrimental to
  survival. Infact survival is optimal when the network has
  intermediate net positive connection strengths. So counter
  intuitively, if positive interactions such as mutualistic or
  symbiotic links dominate other kinds of interactions too much, its
  effect ceases to be beneficial, causing the total population to
  reduce.

\subsection*{Local Dynamics}

Now we attempt to correlate these global features to local
species-level dynamics. Namely, we attempt to correlate the survival
and global stability of the ecological network to dynamical patterns
emerging in the network as a result of interactions.

From the bifurcation diagrams displayed in Fig. 4, one can clearly
discern the presence of coherent collective dynamics in the
system. This coherence breaks down as one approches $\mu=0$, as
evident in Fig. 4, with the network displaying unsynchronized
spatio-temporal chaos.

%%change
In order to gauge the degree of synchronization among the nodes
quantitatively we calculate the synchronization order parameter
$Z_{sync}$. Our attempt now will be to find the {\em correlation
  between synchronization and survival}.  This is an important
question, as synchronization has often been seen as increasing risks
of extinction.
%%Change
Fig. 5 exhibits this synchronization error, along side the number of
active patches (i.e. nodes with non-zero population), the total
population and collective dynamics of the whole network.  It is clear
that {\em synchronization does not necessarily lead to extinction.} In
fact for positive mean interactive strengths, even when the entire
system is completely synchronized, {\em all nodes are active}. The
rationale for the above observation is that synchronization is a
threat only when the synchronized dynamics covers a large range of
population densities, such as in synchronized chaos, which typically
is ergodic over state space. Here on the other hand, the synchronized
dynamics is confined to the ``safe zone'' and the attractor trajectory
does not enter the extinction region \cite{epjb}. So the synchronized
patches survive.

We further investigate the nature of the time series of the local
patches to discern cluster formation, and the phase relation between
the clusters. We find that when the mean of the interaction strengths
has a low positive value, the populations are attracted to a period
$2$ cycle, and the sytem divides into two anti-phase clusters
(cf. Fig.~6). Namely, alternately in time, one set of nodes in the
network have low population densities, while the other set has high
population densities. This behaviour is reminiscent of the field
experiment conducted by Scheffer et al \cite{self-perpetuating} which
showed the existence of self-perpetuating stable states alternating
between blue-green alage and green algae.

%%change
We also studied the phase clusters emerging in the system, by
calculating a phase order parameter $Z_{phase}$, which gives the
fraction of species whose dynamics are in-phase in the network. This
quantity is $\sim 0.5$ in a large range of positive mean interaction
strengths ($\mu \sim 0.1 - 0.6$), indicating that here the network
always splits into two approximately equal clusters. In each cluster
the nodes are in-phase with respect to each one another, and
anti-phase with respect to nodes in the other cluster. However note
that the degree of complete synchronization, which depends on both
phase and amplitude, will be dependent on $\mu$ (as evident from
Fig. 5). So changing the mean interaction strength changes the nature
of the dynamics without destroying this phase relationship and two
phase synchronized clusters of varying amplitudes emerge in a large
range of $\mu$.

\subsection*{Effective map for nodal dynamics}

%%Change
To gain further understanding of the dynamical patterns, we construct
an effective map to mimick the essential dynamics of the nodes. Our
approach is to split the interactive part in Eqn. 2 into an average
part and a term capturing the fluctuations. Here the mean interaction
strength, which is the dominant contribution, is $\mu C$, as there are
a fraction $C$ of non-zero interaction strengths drawn from a
distribution with mean $\mu$. So as first approximation, neglecting
fluctuations, we can model the local dynamics as:

%\begin{equation} 
%X (n+1) = \left\{ \begin{array}{rl}f({X}(n)) + \gamma X(n)  
%& \ \ \ \ \mbox{
%	        if $x > x_{threshold}$,} \\ 0 & \ \ \ \ \mbox{ otherwise,} \end{array} \right.
%\label{effective_map}
%\end{equation}	

%%change
% \begin{equation}
% X(n+1) = f({X}(n)) + \mu C \ X(n)  
% \label{effective_map}
% \end{equation}
\begin{equation}
X(n+1) = f({X}(n)) + \mu C \ X(n)  
\label{effective_map}
\end{equation}
when $X (n+1) > x_{threshold}$, and $X (n+1) = 0$ otherwise. Such an
effective map is an accurate representation of the population dynamics
when there is a high degree of coherence in the system.

To further gauge if the bifurcation diagram obtained numerically in
Fig. 4(d) can be understood qualitatively using this effective map
picture, we argue that the deviations from the effective dynamics can
be modelled by random fluctuations about the mean interaction strength
$\mu C$. So we study the dynamics given by Eqn.~\ref{effective_map}
under the influence of multiplicative noise as well. The results are shown
in Fig. 7. It can be clearly seen that the effective dynamical map,
under random noise, qualitatively captures the collective dynamics of
the multi-species communities.

{\em Analysis:} We can also straight-forwardly analyse the effective
map dynamics given by Eqn.~\ref{effective_map} to find the windows
where the positive steady state is stable. Note that this non-trivial
fixed point, which is a function of $\mu C$, can be obtained as a
solution of:
\begin{equation}
X^{*} = \frac{1}{1- \mu C} f(X^{*})
\label{fp}
\end{equation}
and its stability is determined by the condition $ |f^{\prime}(X^{*})
+ \mu C| < 1$.  Clearly as $\mu C \rightarrow 1$, the fixed point
becomes unboundedly large. This also explains the presence of the
critical positive mean $\mu_c^{explosion}$, with $\mu_c^{explosion} C
\sim 1$, in the system.  From Fig. 8A it is clear that the
  parameter yielding fixed populations in the system is very close
  that found in the effective map (cf. Fig. 7).

Further we employed another, more accurate, approach to gauge the
stability of the synchronized steady state. In this extension, the
stability analysis takes into account the entire network by
considering the extremal eigenvalues of the connectivity matrix. This
yields stability conditions on the fixed point (which is a solution of
Eqn.~\ref{fp}) given by:
\begin{equation}
  f^{\prime}(X^{*}) + \lambda_{max} < 1, \hspace{1in} f^{\prime}(X^{*}) + \lambda_{min} > -1
\label{bounds}
\end{equation}
where, $\lambda_{max}$ and $\lambda_{min}$ are the average maximum and
minimum eigenvalues of the random gaussian matrix, ${\bf J}$.

From Fig. 8B, one observes that the region where synchronized steady
state is stable, namely where $f^{\prime}(X^{*})$ lies within the two
bounds, corresponds quite closely to $\mu \sim 0.8$. This matches the
value observed in simulations very closely (cf. Fig.4), and thus
provides an accurate description of the effective collective behavior
of the system.

%%change
Lastly, consider the scenario that leads the dynamics of the nodes to
extinction, namely to $X^* = 0$. This will happen if $X(n+1) <
x_{threshold}$, which then will map to $X=0$. Notice that for
populations very close to zero, $f(X) \sim 0$.  So from
Eqn. \ref{effective_map}, $X(n+1) \sim \mu C$. This implies that the
subsequent iterate can become negative if and only if $\mu$ is
negative, as $C$ is non-negative and $f(X) > 0$ if $X > 0$. This
suggests why extinctions are seen to arise for $\mu < 0$.

\subsection*{Effect of the degree of variability in inter-species interactions}

Having gained understanding of the collective dynamics of the system
in terms of the dynamics of a single effective stochastic map, we now
try to understand the effect of the standard deviation $\sigma$ of the
connectivity matrix $J$ on the dynamics of the multi-species
community. It seems reasonable to argue that the strength of the noise
term in the effect dynamical map is directly related to $\sigma$,
namely we can associate the stochasticity in the effective map to
variability in the inter-species interaction strength across the
network. Thus we investigate the changes in the bifurcation structure
% for three different values of $\sigma$, which represents different
%%change
for two different values of $\sigma$, which represents different
degrees of spatial variability in the network. From Fig. 9, one can
observe that with increasing $\sigma$, the bifurcation diagram gets
more noisy. This indicates that one can incorporate the spatial
variability in interaction structure easily in the effective map.

%%change
Also, as discussed earlier, for populations close to zero,
Eqn. \ref{effective_map} effectively gives $X(n+1) \sim \mu C$, and
these are driven to extinction if $X(n+1) < x_{threshold}$. This
implies that the sign of the subsequent iterate for a system close to
zero is very sensitive to large fluctuations in the distribution of
interaction strengths. Namely, large variability around the mean value
$\mu C$ in Eqn. \ref{effective_map} can push the system into the
extinction zone, or out of it, when $\mu$, $C$ and $X$ are close to
zero. This accounts for the spread in $\langle N_{active} \rangle$
values around $\mu \sim 0$, in the presence of large $\sigma$ in
Fig. 1.

%   In order to find the origins of fluctuations in the bifurcation
% diagram, we investigate how bifurcation structure changes with
% $\sigma$ as shown in Fig. 10.

\section*{Discussions}

In summary, we have analyzed the survival properties in ecological
networks. In particular, we considered a complex network of
populations where the links are given by a random asymmetric
connectivity matrix $J$, with fraction $1-C$ of zero entries, where
$C$ reflects the over-all connectivity of the system. The non-zero
elements are drawn from a gaussian distribution with mean $\mu$ and
standard deviation $\sigma$. The signs of the elements $J_{ij}$ of the
connectivity matrix $J$ reflect the nature of density-dependent
interactions, such as predatory-prey, mutualism or competition, and
their magnitude reflect the strength of the interaction. Unlike many
earlier studies, we investigate the full range of mean interaction
strengths, and do not confine ourselves to the balanced situation
which assumes $\mu=0$.

Also note that one can potentially draw a parallel between our model
and the system of metapopulations with {\em density dependent
  dispersal} \cite{metapopulation_review}. Namely, our system can also
be interpreted as a network of metapopulation patches \cite{levin1},
or ``a population of populations'' \cite{levin2}. In particular, it
can describe a system comprising many spatially discrete
sub-populations connected by migrations where inter-patch dispersal is
both high enough to ensure demographic connectivity among patches, yet
low enough to maintain some degree of independence in local population
dynamics. The connectivity matrix in this scenario reflects density
dependent dispersal and migration, as is commonly seen in vertebrate
and invertebrate populations
\cite{dispersal1,dispersal2,dispersal3,dispersal4,dispersal5,dispersal6}.

A problem of vital importance here is understanding how broad
features, such as the connectedness and net positive interaction
strength, modulates the emergent dynamics in such a network. First, in
order to gauge the global stability of the system, we calculate the
average number of active nodes, namely number of non-extinct species,
in the network. We find that the network transitions from a completely
extinct system to one where all nodes are active, as the mean
interaction strength goes from negative to positive. This transition,
marked distinctly by scaling relations, gets sharper with increasing
$C$ and decreasing $\sigma$. This result has much relevance, as
realistic ecosystems are unlikely to have a perfect balance of
interactions. So understanding the implications of imbalance in
interaction types and strengths in the network (namely $\mu \ne 0$) is
important.

Another significant observation is that the total population,
reflecting the biomass production in a multi-species community,
displays distinct non-monotonic scaling behaviour with respect to the
product $\mu C$, implying that survival is dependent not merely on the
number of links, but rather on the combination of the sparseness of
the connectivity matrix and the net interaction strength.
Interestingly, in an intermediate window of positive $\mu C$, the
total population is maximal, indicating that too little or too much
positive interactions is detrimental to survival. Infact survival is
optimal when the network has intermediate net positive connection
strengths. Counter-intuitively then, if positive interactions such as
mutualistic or symbiotic links are too dominant, its effect ceases to
be beneficial and in fact results in reduction of the total population.

At the local level we observed marked qualitative changes in dynamical
patterns, ranging from fixed points to spatioteporal chaos, under
variation of mean $\mu$ of the interaction strengths. Specifically we
found anti-phase clusters of period $2$ cycles and the presence of
period-$2$ chaotic bands, in certain windows of mean $\mu$. This
behaviour is reminiscent of the field experiment conducted by Scheffer
et al \cite{self-perpetuating} which showed the existence of
self-perpetuating stable states alternating between blue-green alage
and green algae.  We also studied the correlation between
synchronization and survival, and find that synchronization does not
necessarily lead to extinction.  Lastly, we proposed an effective low
dimensional map to capture the behavior of the entire network, and
this provides a broad understanding of the interplay of the local
dynamical patterns and global stability of the network.

\begin{figure}[ht]
\begin{center}
\includegraphics[width=0.8\linewidth,angle=270]{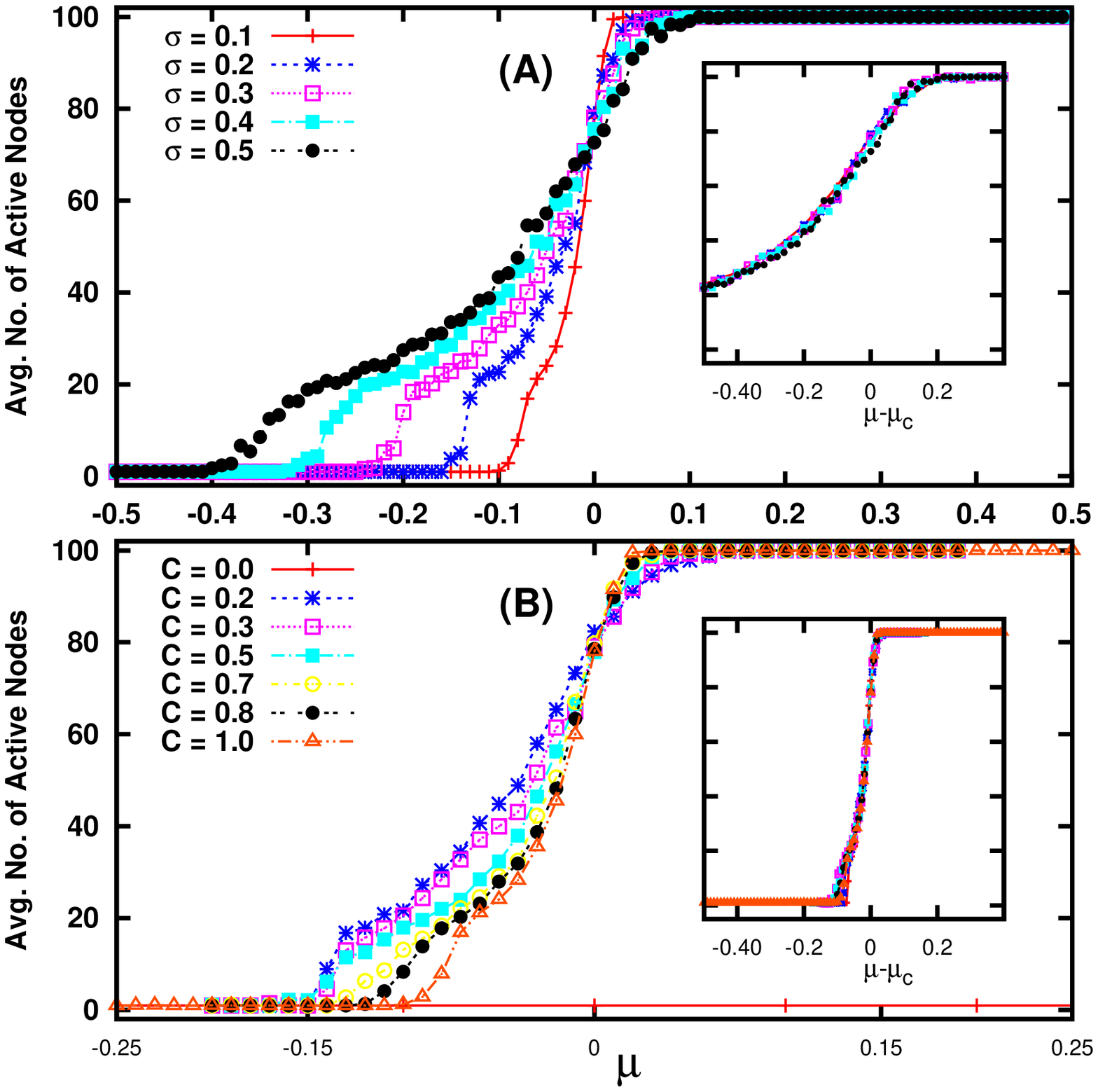}
%{Fig1.eps}
\label{fig1}
\caption{Average number of active nodes, $\langle N_{active} \rangle$,
  as a function of mean $\mu$ of the interaction strengths of the
  connectivity matrix $J$, for different values of connectivity $C$
  and different variabilities in interaction strengths $\sigma$. Here
  system size $N = 100$, and so $\langle N_{active} \rangle = N = 100$
  in the figures reflect a network where no species becomes extinct.
  The insets show the collapse of the scaled $\langle N_{active}
  \rangle$, with respect to $\mu - \mu_c$, where critical $\mu_c =
  0$.}
\end{center}
\end{figure}

\begin{figure}[ht]
\begin{center}
\includegraphics[width=0.8\linewidth,angle=270]{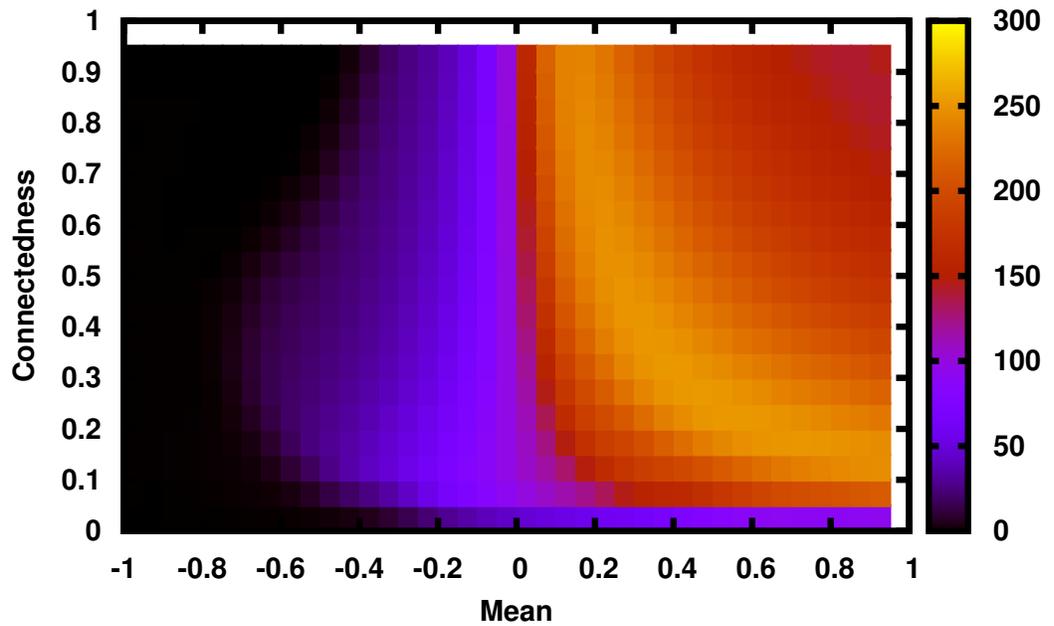}
\label{fig2}
\caption{ Average total population $\langle x_{total} \rangle$ of a
  system of size $N=100$ (represented by the color scale), as a
  function of mean $\mu$ of the interaction strengths and the
  connectedness $C$ (giving the number of non-zero entries in
  connectivity matrix $J$). Here standard deviation $\sigma$ of the
  non-zero matrix elements is $0.5$.}
\end{center}
\end{figure}

\begin{figure}[ht]
\begin{center}
\includegraphics[width=0.8\linewidth,angle=270]{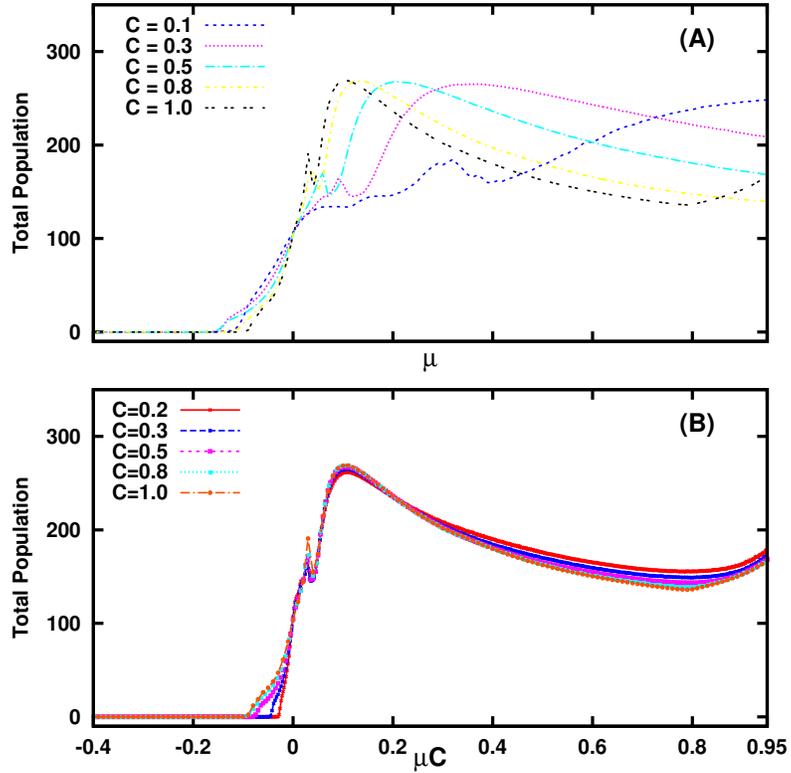}
\label{fig3}
\caption{Average total population of $\langle x_{total} \rangle$ of a
  system of size $N=100$, as a function of mean $\mu$ of the
  interaction strengths for (A) different values of connectedness
  parameter $C$.  In (B), the average total population is given as a
  function of the product $\mu \ C$, and it is evident that the data
  collapses onto a single curve implying a functional relation between
  total population and $\mu \ C$.  Here the average total population
  is obtained by averaging the total over $100$ random realizations,
  and the standard deviation $\sigma$ of the connectivity matrix $J$
  is $0.1$. }
\end{center}
\end{figure}

\begin{figure}[ht]
\begin{center}
\includegraphics[width=0.8\linewidth,angle=270]{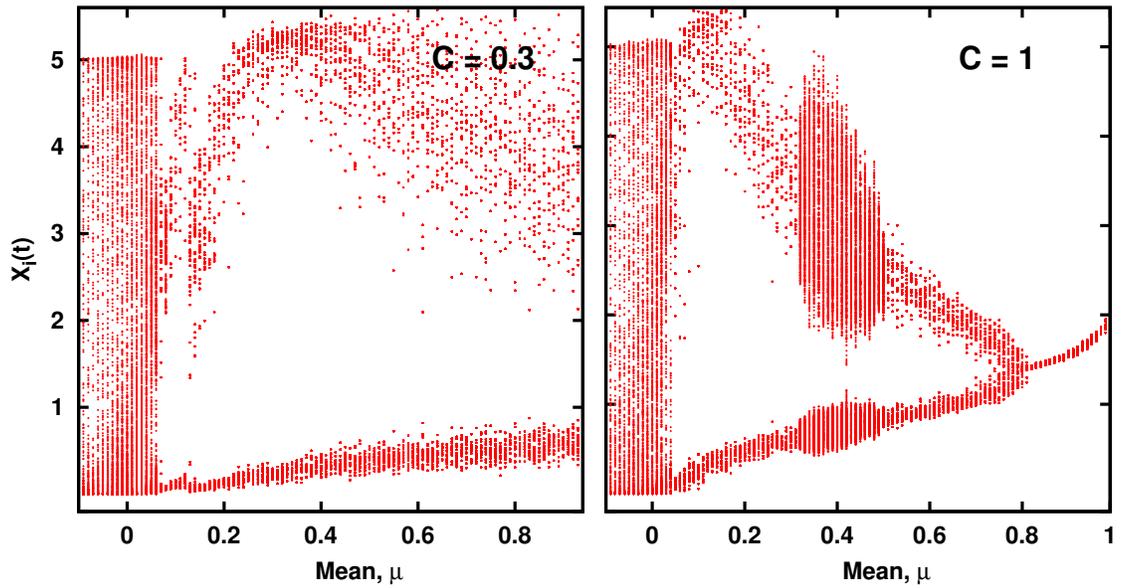}
\label{fig4}
\caption{ Bifurcation Diagram of the population dynamics of the network 
of $100$ species, as a function of mean $\mu$ of the interaction strengths, for representative values of connectedness: (a) $C=0.3$ and (b) $C=1$.
%(b) $C=0.5$, (c) $C=0.8$, (d) $C=1$.
%%Change
Here we show $x_i (t)$, for all $i=1, \dots N$, over a period of time,
after transience. Standard deviation $\sigma$ of $J$ is $0.1$. On
careful observation we found that all bifurcation diagrams collapse on
to a single pattern when viewed as a function of $\mu C$.  }
\end{center}
\end{figure}

\begin{figure}[ht]
\begin{center}
\includegraphics[width=0.8\linewidth,angle=270]{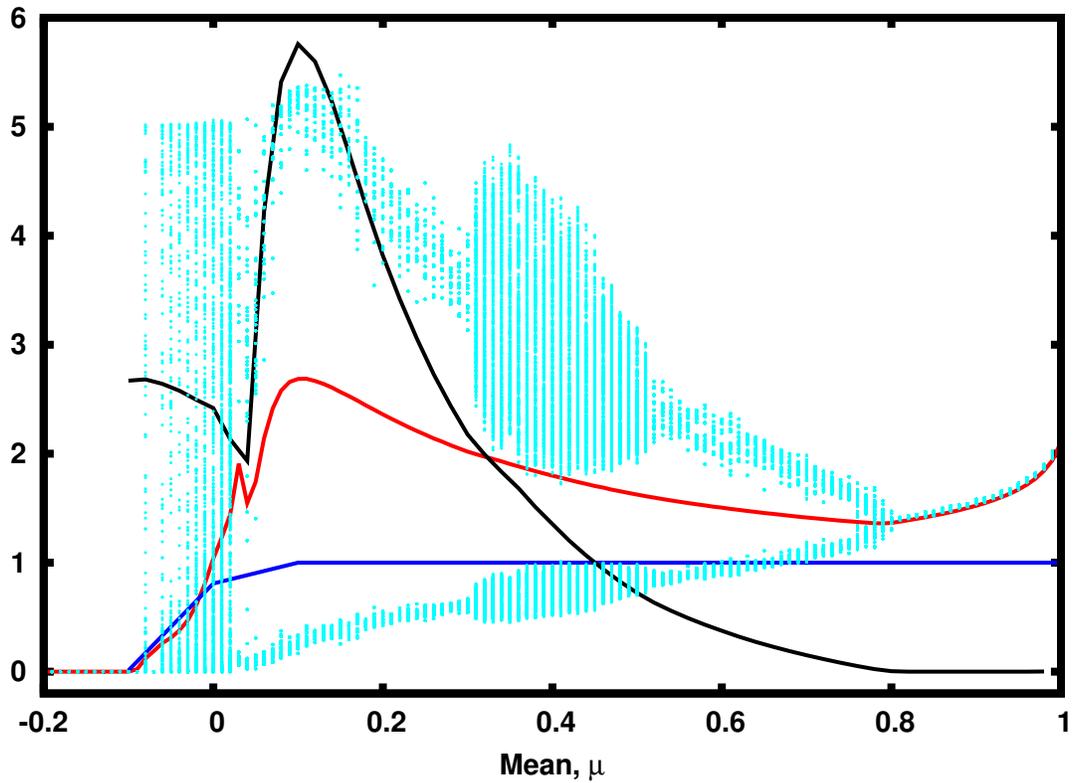}
\label{fig5}
\caption{Synchronization error (dotted black), total population (red),
  bifurcation diagram for the emergent behavior of the population
  densities (cyan) and the average number of active nodes (blue), as a
  function of mean $\mu$ of the interaction strengths. Here the
  connectivity matrix $J$ has $C=1$, $\sigma = 0.1$ and the size of
  network is $N = 100$.  }
\end{center}
\end{figure}

\begin{figure}[ht]
\begin{center}
\includegraphics[width=0.8\linewidth,angle=270]{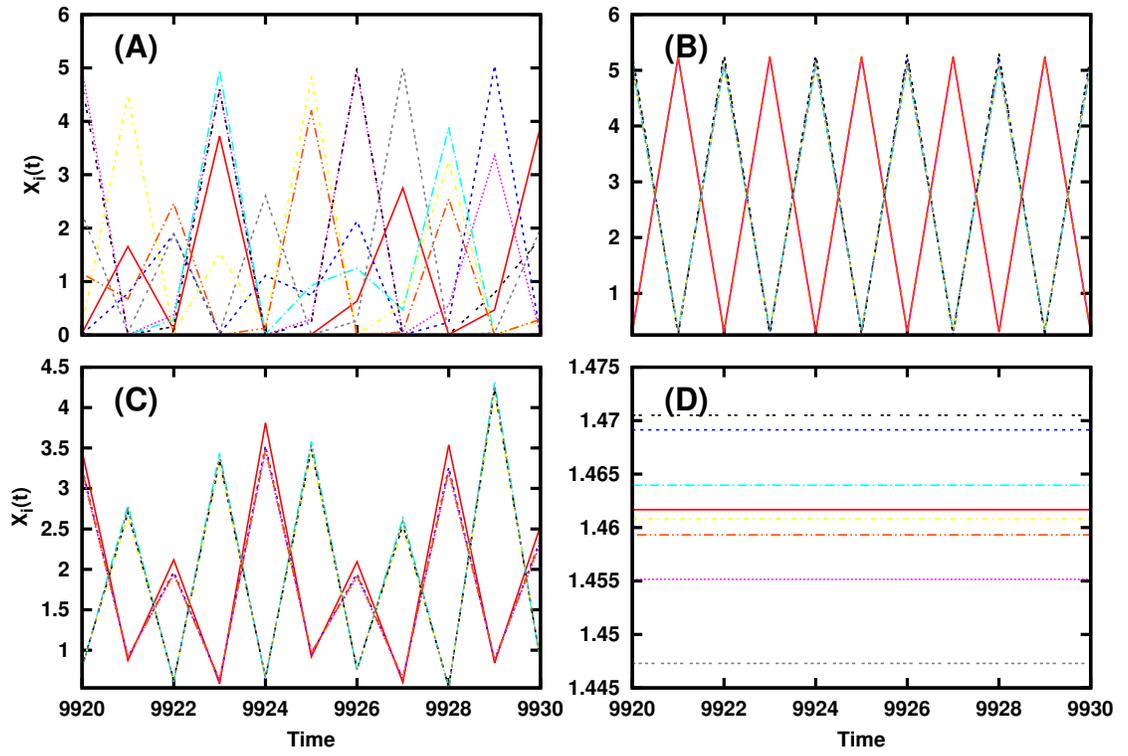}
\label{fig6}
\caption{ Time series of $10$ representative sites in a network of
  $100$ species, for different mean $\mu$ of the interaction
  strengths: (a) $\mu=0$ (b) $\mu = 0.12$ (c) $\mu= 0.40$ and (d) $\mu
  = 0.85$. Here the connectivity matrix $J$ has $C=1$ and $\sigma =
  0.1$.  }
\end{center}
\end{figure}

\begin{figure}[ht]
\begin{center}
\includegraphics[width=0.8\linewidth,angle=270]{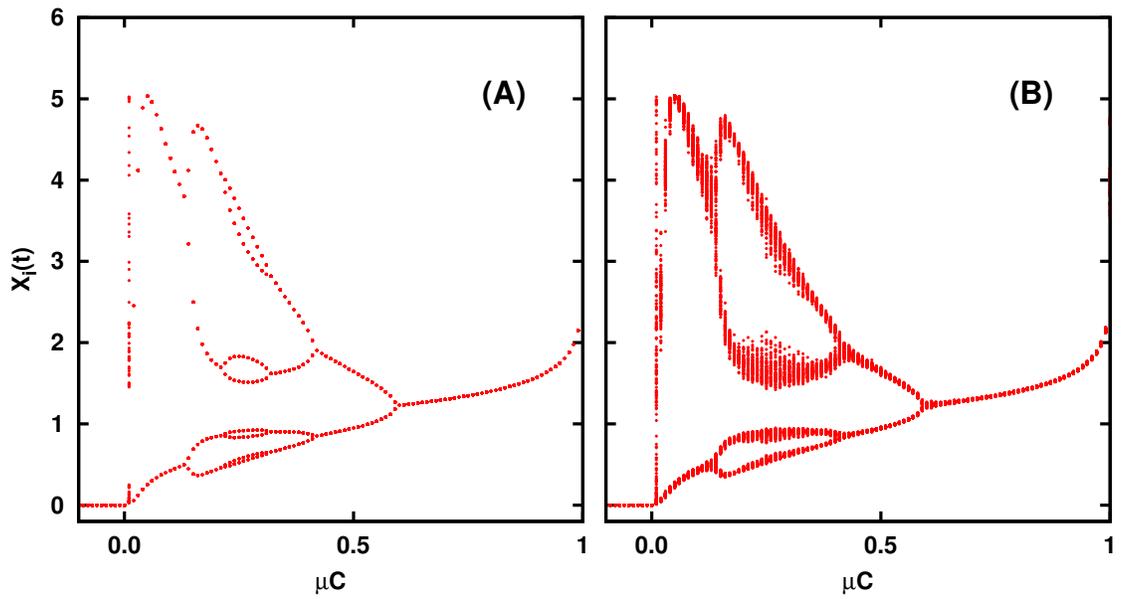}
\label{fig7}
\caption{ Bifurcation diagrams of (Left) the effective map given by
  Eqn.~\ref{effective_map}, as a function of $\mu C$ and (Right) the
  same map under fluctuations given as: $X(n+1) = f({X}(n)) + (\mu C +
  D\xi ) \ X(n)$ , where $\xi$ is a random variable drawn from a
  zero-mean gaussian distribution and $D$ governs the strength of
  fluctuations. Here $D=0.0003$.}
\end{center}
\end{figure}

\begin{figure}[ht]
\begin{center}
\includegraphics[width=0.8\linewidth,angle=270]{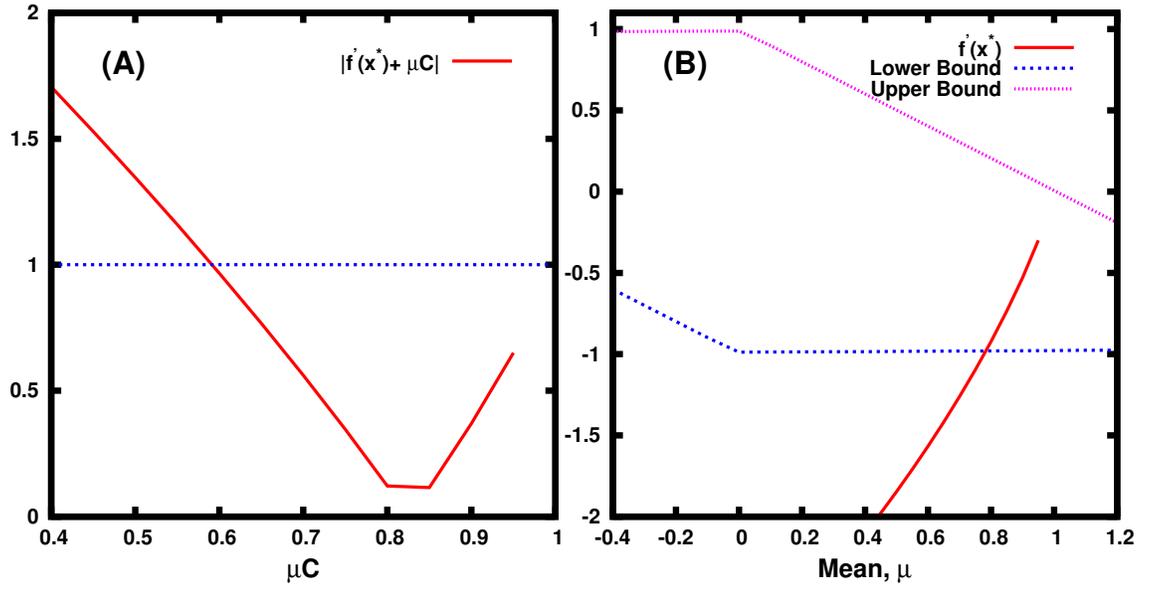}
\label{fig8}
\caption{(A) Stability curve for the fixed point obtained from
Eqn.~\ref{fp}. The fixed point loses stability when $
|f^{\prime}(x^{*}) + \mu C| > 1$, and this occurs for $\mu C \sim
0.6$, which is consistent with the results in Fig. 7(B) Upper bound
($1-\lambda_{max}$), lower bound ($-1-\lambda_{min}$) and
$f^{\prime}(x^{*})$. The region where the $f^{\prime}(x^{*})$ lies
between upper bound and lower bound (cf. Eqn. \ref{bounds}) gives us
the region of stability of global synchronized state of the network.
}
\end{center}
\end{figure}

\begin{figure}[ht]
\begin{center}
\includegraphics[width=0.8\linewidth,angle=270]{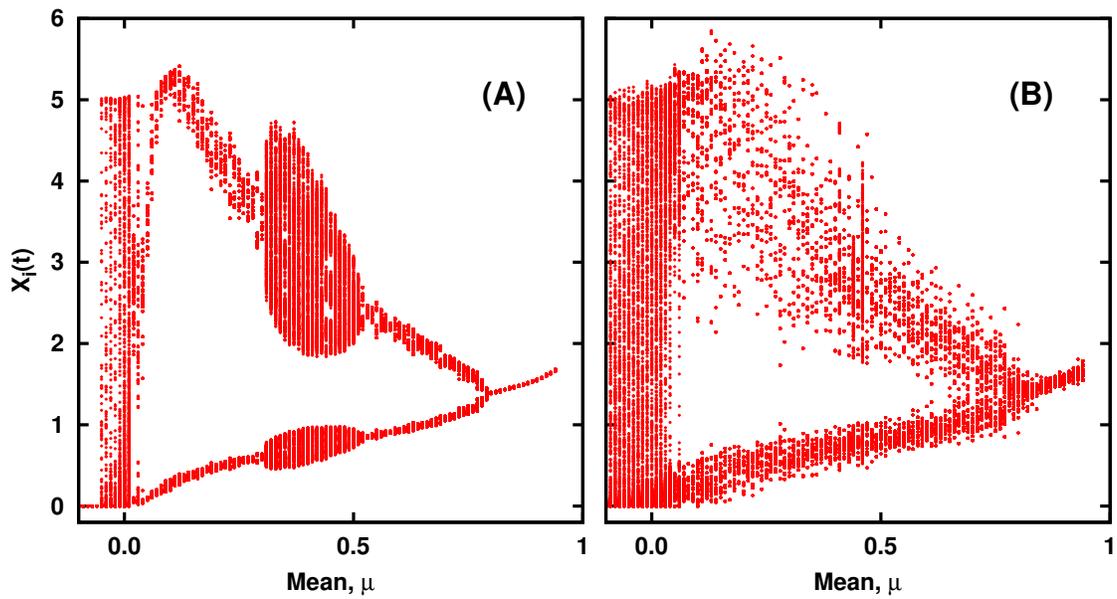}
\label{fig9}
\caption{Bifurcation diagram of the population dynamics of all
the species as mean interaction strength $\mu$ is varied, for two
different values of standard deviation: (a) $\sigma = 0.05$ and (b)
$\sigma = 0.5$. For this representative case, the network is taken to
be fully connected, i.e. $C=1$.
}
\end{center}
\end{figure}

\end{document}